\documentclass[conference]{IEEEtran}

\usepackage{graphicx}
\usepackage{hhline}
\usepackage{times}
\usepackage{epsfig}
\usepackage{colortbl}
\usepackage{float}
\usepackage{subfig}
\usepackage{verbatim}

\usepackage[latin1]{inputenc}
\usepackage[T1]{fontenc}
\usepackage{lmodern}
\usepackage{amsmath}
\usepackage{amsthm}
\usepackage{amssymb}
\usepackage{bbm}

\usepackage{tikz}
\usepackage{pgfplots}
\usepackage[right]{eurosym}
\usepackage{multirow}
\usepackage{color}
\usepackage{psfrag}
\usepackage{todonotes}
\usepackage[normalem]{ulem} 
\usepackage{cancel}
\usepackage{cite} 

\usepackage{cleveref} 
\usepackage{pdfpages} 

\newtheorem{theorem}{Theorem}
\newtheorem{lemma}[theorem]{Lemma}
\newtheorem{definition}[theorem]{Definition}

\DeclareGraphicsExtensions{.pdf} 

\IEEEoverridecommandlockouts
\begin{document}

\title{Joint secrecy over the $K$-Transmitter Multiple Access Channel
\thanks{This work is supported in part by DFG Grant CH 601/2-1 and NSF award CNS-1617335.}
}

\author{\IEEEauthorblockN{Yanling Chen\IEEEauthorrefmark{1}, 
O.~Ozan~Koyluoglu\IEEEauthorrefmark{2} and 
 A. J. Han Vinck\IEEEauthorrefmark{1}
 }
 
\IEEEauthorblockA{\IEEEauthorrefmark{1}
Institute of Digital Signal Processing, University of Duisburg-Essen, Germany.\\
}

\IEEEauthorblockA{\IEEEauthorrefmark{2}
Department of Electrical and Computer Engineering,
The University of Arizona.\\
Email: \{yanling.chen, han.vinck\}@uni-due.de, ozan@email.arizona.edu.}
}

\maketitle

\begin{abstract}
This paper studies the problem of secure communication over a $K$-transmitter multiple access channel in the presence of an external eavesdropper, subject to a joint secrecy constraint (i.e., information leakage rate from the collection of $K$ messages to an eavesdropper is made vanishing). As a result, we establish the joint secrecy achievable rate region. To this end, our results build upon two techniques in addition to the standard information-theoretic methods. The first is a generalization of Chia-El Gamal's lemma on entropy bound for a set of codewords given partial information. The second is to utilize a compact representation of a list of sets that, together with properties of mutual information, leads to an efficient Fourier-Motzkin elimination. These two approaches could also be of independent interests in other contexts.
\end{abstract}
\section{Introduction}
For the problem of reliably communicating independent messages over a multiple access channel (MAC), Ahlswede \cite{src:Ahlswede1971} first studied the $2$-transmitter and $3$-transmitter cases and determined the respective capacity regions; whilst Liao \cite{src:Liao1972MAC} considered the general $K$-transmitter MAC and fully characterized its capacity region. 

Inspired by the pioneering works of Wyner \cite{src:Wyner1975} and Csisz{\'a}r and K\"{o}rner \cite{Csisz'ar:Broadcast78} that studied the information theoretic secrecy of a point-to-point communication in the presence of an external eavesdropper, a MAC with an external eavesdropper was first introduced in \cite{src:TekinYener2008}. In particular,  \cite{src:TekinYener2008} focused on a degraded Gaussian MAC with $K$-transmitters and established achievable rate regions subject to a pre-specified secrecy measure; while a discrete memoryless 2-transmitter MAC with an external eavesdropper was considered in \cite{src:TangLiu2007}. Further works on MAC with an external eavesdropper include but not limited to \cite{src:TekinYenerMACTW2008, src:Elrem2008, Koyluoglu:Cooperative11, src:Wiese2013}. However, its secrecy capacity region, even for the 2-transmitter case, still remains an open problem.

\begin{figure}[tb]
 \centering
  \includegraphics[width=0.45\textwidth]{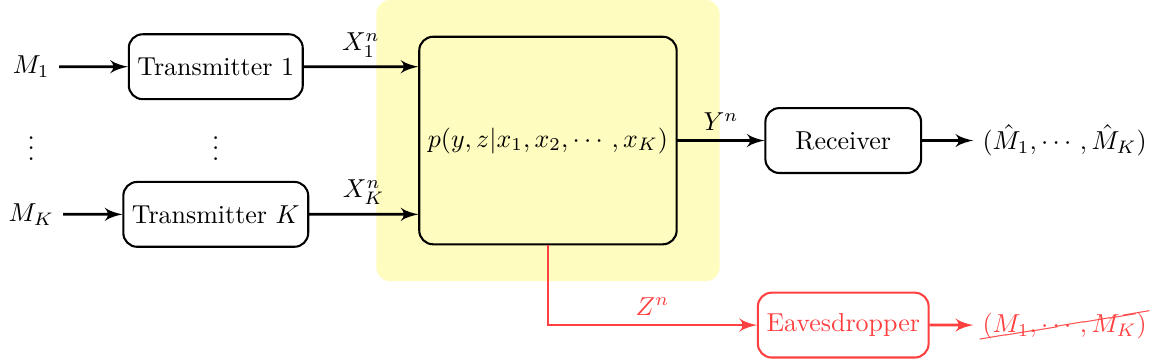}
  \caption{$K$-transmitter DM-MAC with an eavesdropper.} \label{fig: 3Tx DM-MAC with an external eavesdropper}
 \end{figure}

In this paper, we consider the secure communication over a $K$-transmitter MAC subject to the joint secrecy constraint (i.e., information leakage rate from the collection of $K$ messages to an eavesdropper is made vanishing), the channel model of which is shown in Fig. \ref{fig: 3Tx DM-MAC with an external eavesdropper}. As a general result, we establish a joint secrecy achievable rate region. 

The rest of the paper is organized as follows. Section~\ref{sec:model} introduces the system model and presents necessary definitions; Section~\ref{sec: main result1} gives the general conditional entropy bound. The bound plays an important role in establishing the joint secrecy rate region that is given in Section \ref{sec: main result1}. To enhance the flow of the paper, some details for the Fourier-Motzkin elimination are relegated to Appendix \ref{App: Compact form}. 

\section{Preliminaries}\label{sec:model}

Consider a discrete memoryless MAC (DM-MAC) with $K$ transmitters, one legitimate receiver, and one passive eavesdropper, which is defined by $p(y,z|x_1,x_2, \cdots, x_{K}).$ The transmitter $i,$ aims to send message $m_i,$ to the legitimate receiver, where $i\in \mathcal{K}=\{1,2,\cdots, K\}$. Suppose that $x_i^n$ is the channel input at transmitter $i$, and the channel outputs at the legitimate receiver and eavesdropper are $y^n$ and $z^n$, respectively. By the \emph{discrete memoryless} nature of the channel (without any feedback), we have
\begin{equation} \label{def: dm}
	p(y^n,z^n|x_1^n,\cdots, x_K^n) = \prod_{i=1}^{n} p(y_{i}, z_{i}|x_{1,i},\cdots, x_{K,i}). 
\end{equation}

A $(2^{nR_1}, 2^{nR_2},\cdots, 2^{nR_{K}}, n)$ secrecy code $\mathcal{C}_n$ for the DM-MAC consists of
\begin{itemize}
	\item $K$ message sets $\mathcal{M}_1, \mathcal{M}_2, \cdots, \mathcal{M}_{K},$ where $m_i\in \mathcal{M}_i=[1:2^{nR_i}]$ for $i\in\mathcal{K};$
	\item $K$ encoders each assigning a codeword $x_i^n$ to message $m_i$ for $i\in\mathcal{K};$ and 
	\item One decoder at the legitimate receiver that declares an estimate of $(m_{1},m_{2}, \cdots, m_{K})$ say $(\hat{m}_{1},\hat{m}_{2}, \cdots, \hat{m}_{K})$ or an error to the received sequence $y^n.$ 
\end{itemize}
 
In this paper, for a list of random variables $W_j$ for $j\in \mathcal{K},$ and a fixed $\mathcal{J}\subseteq\mathcal{K},$ we denote $W_{\mathcal{J}}=\{W_i|i\in \mathcal{J}\}.$ 
Assume that the messages $M_{\mathcal{K}}$ are uniformly distributed over their corresponding message sets. Therefore, we have 
	$R_i =\frac{1}{n}H(M_i),$  for $i\in\mathcal{K}.$
Denote the \emph{average probability of decoding error} at the legitimate receiver as  
$P_{e}^n(\mathcal{C}_n)=\Pr\left\{\bigcup\limits_{i\in \mathcal{K}} \{M_i\neq \hat{M}_i\}|\mathcal{C}_n\right\}.$ 
Define the \emph{joint} information leakage rate to the eavesdropper by $R_{L,{\mathcal{K}}}(\mathcal{C}_n)= \frac{1}{n} I(M_{\mathcal{K}}; Z^n|\mathcal{C}_n).$  
The rate pair $(R_1,R_2,\cdots, R_{K})$ is said to be \emph{achievable under the joint secrecy constraint}, if there exists a sequence of $(2^{nR_1}, 2^{nR_2}, \cdots, 2^{nR_{K}}, n)$ codes $\{\mathcal{C}_n\}$ such that
\begin{align} 
	  P_{e}^n(\mathcal{C}_n) &\leq \epsilon_n, \label{eq:Reliability} \\
	  R_{L,\mathcal{K}}(\mathcal{C}_n)	&\leq	\tau_n,  \label{eq:IndSec} \\
	  \lim\limits_{n\to\infty} \epsilon_n&= 0 \quad \mbox{and} \quad \lim\limits_{n\to\infty} \tau_n= 0. \label{eq:n to infty} 
\end{align}
  
Recall that $\mathcal{K}=\{1,2,\cdots, K\}.$ We have the following definition and lemmas. 
\begin{definition}\label{app: def indicator vector}
	The indicator vector of a subset $\mathcal{T}$ of set $\mathcal{K},$ denoted by $\mathrm{1}_\mathcal{T},$ is a $1\times K$ vector, with its $i$-th element equal to 1 if $i\in\mathcal{T}$ and 0 otherwise, for $1\leq i\leq K.$
\end{definition}
For instance, for $K=5,$ $\mathcal{K}=\{1,2,3,4,5\}$ and $\mathcal{T}=\{1,3,5\},$ we have $\mathrm{1}_\mathcal{T}=[1 \ 0\  1\  0\  1],$ and $\mathrm{1}_\mathcal{\emptyset}=[0 \ 0\  0\  0\  0].$

Let $\{\mathcal{T}_i |1\leq i\leq t\}$ be a list of $t$ subsets of $\mathcal{K}.$ 
\begin{definition}\label{app: def presence vector}
The presence vector of $\{\mathcal{T}_i |1\leq i\leq t\}$ is defined to be  $\mathrm{t}_{\#}=\sum\limits_{i=1}^{t} \mathrm{1}_{\mathcal{T}_{i}},$ which counts the number of presences of each element of $\mathcal{K}$ over $\{\mathcal{T}_i |1\leq i\leq t\}.$
\end{definition}

\begin{definition}\label{app: def compact form}
A compact form of the element rearrangement for $\{\mathcal{T}_i|1\leq i\leq t\}$ is defined to be $\{\mathcal{T}_{t,i}^{\ast}|1\leq i\leq t\},$ where $\mathcal{T}_{t,i}^{\ast}$ contains the elements that present at least $i$ times from all these $t$ subsets, i.e., 
	\begin{align*}
		\mathcal{T}_{t,i}^{\ast} &= \bigcup_{\{j_1, \cdots, j_i\}\subseteq [1:t]} \left(\bigcap_{k=1}^i \mathcal{T}_{j_k}\right).	
	\end{align*}
	Clearly, $\mathcal{T}_{t,t}^{\ast}\subseteq \mathcal{T}_{t,t-1}^{\ast}\subseteq \cdots \subseteq \mathcal{T}_{t,1}^{\ast}.$ And, $\mathcal{T}_{t,i}^{\ast}=\emptyset$ for $i>t_{\max},$ where $t_{\max}$ is the largest element of $\mathrm{t}_{\#}.$ So $\{\mathcal{T}_{t,i}^{\ast}| 1\leq i\leq t_{\max}\}$ is the compact form without the empty sets. 
\end{definition}

\begin{lemma}\label{lem: invariability of presence vector}
$\mathrm{t}_{\#}=\sum\limits_{i=1}^{t} \mathrm{1}_{\mathcal{T}_{i}}=\sum\limits_{i=1}^{t} \mathrm{1}_{\mathcal{T}_{t,i}^*}=\sum\limits_{i=1}^{t_{\max}} \mathrm{1}_{\mathcal{T}_{t,i}^*}.$
\end{lemma}
For instance, for $K=3,$ $\mathcal{K}=\{1,2,3\}$ and $\{\mathcal{T}_i |1\leq i\leq 3\}$ with $\mathcal{T}_1=\{1\}, \mathcal{T}_2=\{1,2\},$ $\mathcal{T}_3=\{2,3\}$. We have $\mathrm{t}_{\#}=(2,2,1), $ ${t}_{\max}=2,$ $\mathcal{T}_{3,1}^{\ast}=\{1,2,3\}, \mathcal{T}_{3,2}^{\ast}=\{1,2\}, \mathcal{T}_{3,3}^{\ast}=\emptyset.$

\begin{lemma}\label{lem: app compact arragement properties}
Given two lists of sets $\{\mathcal{T}_{1i} |1\leq i\leq t_1\}$ and $\{\mathcal{T}_{2i} |1\leq i\leq t_2\},$ elements of which are $t_1$ and $t_2$ subsets of set $\mathcal{K},$ respectively, if they share the same presence sequence $\mathrm{t}_{\#},$ then they also share the same compact form of the element rearrangement without empty sets. 
\end{lemma}

 \section{Conditional entropy bound}\label{sec: main result1}
In this section, we give the conditional entropy bound, which is a generalization of Chia-El Gamal's lemma \cite[Lemma 1]{Chia:Three-receiver12} on entropy bound for a set of conditionally independent codewords given partial information.
  \begin{lemma}\label{lem: Equivocation at Eve}
  	Let $R_{v,i}\geq 0$ for $i\in\mathcal{K},$ $\epsilon>0,$ and $(Q, V_1, \cdots, V_K, Z) \sim p(q)\cdot \prod\limits_{i\in\mathcal{K}}p(v_i|q)\cdot p(z|v_1,\cdots, v_K).$  Let $Q^n$ be a random sequence and each $q^n=(q(1), \cdots, q(n))$ distributed according to $\prod\limits_{t=1}^{n}p(q(t)).$ For $i\in\mathcal{K},$ let $V_i^n(l_i),$ $l_i\in [1:2^{nR_{v,i}}],$ be a set of random sequences that are conditionally independent given $Q^n$ and each $v_i^n=(v_i(1), \cdots, v_i(n))$ distributed according to $\prod\limits_{t=1}^{n}p(v_i(t)|q(t)),$ and let $\mathcal{C}$ be the codebook of $\left(Q^n, V_1^n(1), \cdots, V_K^n(2^{nR_{v,K}})\right).$ Let $L_i$ be the random index of $V_i^n,$ for $i\in\mathcal{K},$  with an arbitrary probability mass function. Then, if $\Pr\{(Q^n, V_1^n(L_1), \cdots, V_K^n(L_K), Z^n)\in \mathcal{T}_{\epsilon}^n (Q, V_1, \cdots, V_K, Z)\}\allowbreak\to 1$ as $n\to \infty$ and 
 	\begin{equation}\label{eqn:Lem Rates constraints on equivocation}
 		\sum_{j\in \mathcal{J}} R_{v, j} \geq I(V_{\mathcal{J}};Z|Q), \quad \forall \mathcal{J}\subseteq \mathcal{K} 
 	\end{equation}
 	there exists a $\delta_n(\epsilon)\to 0$ as $\epsilon\to 0$ and $n\to \infty,$ such that for $n$ sufficiently large, $H(L_1, \cdots, L_K|Z^n,Q^n, \mathcal{C})\leq n\left[\sum\limits_{j\in \mathcal{K}} R_{v,j}-I(V_{\mathcal{K}};Z|Q)\right]+n\delta_n(\epsilon).$ 
  \end{lemma}
\begin{IEEEproof} 
Given $z^n,$ let us define $\mathcal{L}$ as the set of indices $(l_1, l_2, \cdots, l_K)$ such that 
\begin{equation*}
	\left(q^n, v_1^n(l_1), \cdots, v_K^n(l_K), z^n\right) \in \mathcal{T}_{\epsilon}^n(Q, V_1, \cdots, V_K, Z).
\end{equation*}
First we show that the expected size of this list, over all randomly generated codebooks, is upper bounded by
\begin{equation}\label{eqn: E(L)}
	\mathbb{E}(|\mathcal{L}|)\leq 1+\sum_{i=1}^{2^K-1} 2^{n[\mathrm{I}_i+\delta(\epsilon)]},
\end{equation}
where 
$
	\mathrm{I}_i=\sum\limits_{j\in \mathcal{J}_i} R_{v,j}-I(V_{\mathcal{J}_i};Z|V_{\mathcal{J}_i^c},Q).
$
Here, $\{\mathcal{J}_i|i\in[1:2^K-1]\}$ are the $2^K-1$ non-empty subsets of $\mathcal{K},$ and $\mathcal{J}_i^c= \mathcal{K}\backslash\mathcal{J}_i$ for $i\in[1:2^K-1].$ 
Note that 
\begin{align*}
	\mathbb{E}(|\mathcal{L}|)= & \Pr\{(L_1, \cdots, L_{K})\in \mathcal{L}\} \\
	&+ \sum_{(l_1,\cdots, l_{K})\neq (L_1, \cdots, L_{K})} \Pr\{(l_1,\cdots, l_K)\in \mathcal{L}\},
\end{align*}
where $(L_1, \cdots, L_K)$ are the true indices chosen by the sources. Since $\Pr\{(q^n, v_1^n(L_1), \cdots, v_K^n(L_K), z^n)\in \mathcal{T}_{\epsilon}^n (Q, V_1, \cdots, V_K, Z)\}\to 1$ as $n\to \infty,$ the 1st term tends to 1 as $n\to \infty.$ As for the 2nd term, we can distinguish $(2^K-1)$ cases according to the values of $(l_1,l_2,\cdots, l_K).$ More specifically, for each $\mathcal{J}_i,$ $i\in [1:2^K-1],$ we consider the following case:
\begin{itemize}
	\item $l_j\neq L_j$ for $j\in \mathcal{J}_i,$ and $l_j= L_j$ for $j\in \mathcal{J}_i^c.$ 
	
	In this case, in total there are at most $2^{n\sum\limits_{j\in\mathcal{J}_i}R_{v,j}}$ possible $(l_1,\cdots, l_{K}).$ By the joint typicality lemma, we can upper bound $\Pr\{(l_1,\cdots, l_K)\in \mathcal{L}\}$ by
		\begin{equation*}
			 2^{-nI(V_{\mathcal{J}_i}; V_{\mathcal{J}_i^c}, Z|Q)+n\delta(\epsilon)}\stackrel{(a)}{=} 2^{-nI(V_{\mathcal{J}_i}; Z| V_{{\mathcal{J}_i}^c},Q)+n\delta(\epsilon)},
		\end{equation*} 
		where $(a)$ is due to the fact that $V_{\mathcal{J}_i}$ and $V_{\mathcal{J}_i^c}$ are conditionally independent given $Q.$ 
		
		Therefore, in this case, there are at most $2^{n[\mathrm{I}_i+\delta(\epsilon)]}$ number of $(l_1,l_2,\cdots, l_K)$ falling in the list $\mathcal{L}.$
\end{itemize}
Summing up all the numbers of $(l_1,l_2,\cdots, l_K)$ falling in the list $\mathcal{L}$ over all these $(2^K-1)$ cases, we prove \eqref{eqn: E(L)}.

Furthermore, define the indicator variable $E=1$ if $(L_1, \cdots, L_K) \in \mathcal{L},$ and $E=0$ otherwise. We have
\allowdisplaybreaks
\begin{align*}
	H(L_1,& \cdots, L_K|Z^n,Q^n, \mathcal{C})\leq  H(L_1, \cdots, L_K, E|Z^n,Q^n, \mathcal{C})\\
		\leq & H(E)+ H(L_1, \cdots, L_K|Z^n,Q^n,E,\mathcal{C})\\
		\stackrel{(b)}{\leq} & 1+ H(L_1, \cdots, L_K|Z^n,Q^n,E=1,\mathcal{C})\\
			 &+ \Pr\{E=0\} H(L_1, \cdots, L_K|\mathcal{C}),
\end{align*}
where $(b)$ follows from the fact that $H(E)\leq 1$ since $E$ is a binary random variable;  $\Pr\{E=1\}\leq 1$ and conditioning does not increase the entropy. 

Since $\Pr\{(q^n, v_1^n(L_1), \cdots, v_K^n(L_K), z^n)\in \mathcal{T}_{\epsilon}^n (Q, V_1, \cdots, \allowbreak V_K, Z)\}\allowbreak\to 1$ as $n\to \infty,$ then $\Pr\{E=0\}=\Pr\{(L_1, L_2, \cdots, L_K)\notin \mathcal{L}\}$ can be made arbitrarily small as $n\to \infty.$   Next, 
\allowdisplaybreaks
\begin{align*}
	 H&(L_1, \cdots, L_K|Z^n,Q^n,E=1,\mathcal{C})\\
		\stackrel{(c)}{=} & H(L_1, \cdots, L_K|Z^n,Q^n,E=1,\mathcal{C}, \mathcal{L}, |\mathcal{L}|) \\
		\leq  &H(L_1, \cdots, L_K|E=1, \mathcal{L}, |\mathcal{L}|) \\
		= & \sum_{l\in \mathrm{supp}(|\mathcal{L}|)} \Pr\{|\mathcal{L}|=l\} H(L_1,  \cdots, L_K|E=1, \mathcal{L}, |\mathcal{L}|=l) \\
		\stackrel{(d)}{\leq}& \sum_{l\in \mathrm{supp}(|\mathcal{L}|)} \Pr\{|\mathcal{L}|=l\} \log_2(l) \\
		= & \mathbb{E} (\log_2(|\mathcal{L}|))
		\stackrel{(e)}{\leq} \log_2\left(\mathbb{E}(|\mathcal{L}|)\right)\\
		\stackrel{(f)}{\leq}& n\max\{0, \max_{i\in[1:2^K-1]} \mathrm{I}_i\}+K+ n\delta(\epsilon)\\
		\stackrel{(g)}{\leq} & n \left[\sum_{j\in \mathcal{K}} R_{v,j}-I(V_{\mathcal{K}};Z|Q)\right]+K+ n\delta(\epsilon),		
\end{align*} 
where $(c)$ follows from the fact that $\mathcal{L}$ and $|\mathcal{L}|$ are functions of the output $Z^n,$ given $\mathcal{C}$ and $Q^n;$ $(d)$ is due to the fact that, knowing $E=1,$ the sent indices $(L_1,L_2,\cdots, L_K)$ belong to the list $\mathcal{L}$ and the uncertainty is upper bounded by the log cardinality of the list; $(e)$ is by Jensen's inequality; $(f)$ is by \eqref{eqn: E(L)} along with an application of the log-sum-exp inequality: 
	$\log_2\left(\sum\limits_{x\in \mathcal{X}} 2^x \right)\leq \max\limits_{x\in \mathcal{X}} x +\log_2\left(|\mathcal{X}|\right);$
and $(g)$ follows if the rates satisfies \eqref{eqn:Lem Rates constraints on equivocation}, i.e.:
$
	\sum\limits_{j\in \mathcal{J}}R_{v,j}\geq I(V_{\mathcal{J}};Z), \ \forall \mathcal{J}\subseteq \mathcal{K}.
$
This, along with previous remarks yields the desired inequality
 (by defining $\delta_n(\epsilon)$ to be the arbitrary small term $\mathcal{O}(\epsilon)+(K+1)/n$).
\end{IEEEproof}

 \section{Achievable rate region}\label{sec: main result2}
In this section, we give an achievable joint secrecy rate region of the $K$-transmitter DM-MAC with an external eavesdropper. This result recovers the joint secrecy result for $K=2$ in \cite[Theorem 2]{src:Chen2016MAC}, which improves \cite[(8)]{src:TangLiu2007} with channel prefixing as demonstrated in \cite{src:Chen2016MAC}.
\begin{theorem}\label{Cor: S_MAC_KS}
An achievable joint secrecy rate region of the $K$-transmitter DM-MAC with an external eavesdropper is given by the union of non-negative rate pairs $(R_1,R_2, \cdots, R_K)$ that are defined by the followings:
 \begin{align*}
	\sum_{j\in \mathcal{J}} R_{j} \leq &  I(V_{\mathcal{J}};Y|V_{\mathcal{J}^c},Q)-I(V_{\mathcal{J}};Z|Q), \quad \forall \mathcal{J}\subseteq \mathcal{K}
 \end{align*}
where the union is over input probability distributions that factor as $p(q)\prod\limits_{i\in \mathcal{K}}p(v_i|q)p(x_i|v_i).$
  \end{theorem}  
\begin{IEEEproof}
 Fix $p(q)$ and $p(v_i|q), p(x_i|v_i)$ for $i\in \mathcal{K}$. Generate a random sequence $q^n$, where $p(q^n)=\prod\limits_{t=1}^n p(q(t))$ with each entry chosen as i.i.d. $p(q)$. The sequence $q^n$ is given to every node in the system. 
 
 {\em Codebook generation:} For $i\in \mathcal{K},$ to construct codebook $\mathcal{C}_i$, randomly generate $2^{n[R_i+R_{i,r}]}$ i.i.d. sequences $v_i^{n}(m_i, m_{i,r})$, with $(m_i, m_{i,r})\in[1:2^{nR_i}]\times[1:2^{nR_{i,r}}]$, each with probability $p(v_i^n|q^n)=\prod\limits_{t=1}^n p(v_i(t)|q(t))$. Every node in the network knows these codebooks. Denote the overall codebook as $\mathcal{C}.$ 
 
 {\em Encoding:} For $i\in \mathcal{K},$ to send message $m_i$, transmitter $i$   randomly and uniformly chooses $m_{i,r}\in [1: 2^{nR_{i,r}}]$ and finds $v_i^n(m_i,m_{i,r})$. Then, given the codeword $v_i^n(m_i,m_{i,r}),$ it generates $x_i^n$ according to $\sum\limits_{t=1}^{n}p(x_i(t)|v_i(t))$ and transmits this sequence to the channel.
 
 {\em Decoding:} The legitimate receiver, upon receiving $y^n$, finds  $v_1^n(\hat{m}_1, \hat{m}_{1,r})$, $v_2^n(\hat{m}_2, \hat{m}_{2,r}),$ ... ,  $v_{K}^n(\hat{m}_{K}, \hat{m}_{K,r})$ such that $(v_1^n(\hat{m}_1, \hat{m}_{1,r}), \allowbreak v_2^n(\hat{m}_2, \hat{m}_{2,r}), \cdots, v_{K}^n(\hat{m}_{K}, \hat{m}_{{K},r}), y^n)$ is jointly typical. 

 {\em Analysis of the error probability of decoding:}  Consider the expected value of the error probability of decoding over the ensemble of random codes $\mathcal{C},$ i.e., $P_{e}=\mathbb{E}\left[P_{e}(\mathcal{C})\right].$ Note that here $\mathcal{C}$ denotes the random variable that represents the randomly generated codebook that adhere to the above scheme. From the decoding analysis for the multiple access channel, see, e.g., \cite{ElGamal:2012}, $P_{e}$ can be made approximately zero as $n\to\infty$ if
 \allowdisplaybreaks
 \begin{align}
 \begin{split} \label{eqn:RateConstraints}
		\sum_{j\in \mathcal{J}} [R_{j}+R_{j,r}] \leq I(V_{\mathcal{J}};Y|V_{\mathcal{J}^c},Q), \quad \forall \mathcal{J}\subseteq \mathcal{K}.
 \end{split}
 \end{align}
 
 {\em Analysis of joint secrecy:} 
 For the joint secrecy as defined in \eqref{eq:IndSec}, we show in the following that $\mathbb{E}\left[R_{L,\mathcal{K}}(\mathcal{C})\right]\leq \tau_n.$ To this end, we show that $H(M_{\mathcal{K}}|Z^n,Q^n, \mathcal{C})\geq n\sum\limits_{i\in \mathcal{K}}R_i-n\tau_n,$
 as this implies $I(M_{\mathcal{K}};Z^n|\mathcal{C})\leq I(M_\mathcal{K};Z^n,Q^n|\mathcal{C})\leq n\tau_n$. 
 \allowdisplaybreaks
 	\begin{align*}
			  H&(M_\mathcal{K}|Z^n,Q^n,\mathcal{C})
		 	= H(M_\mathcal{K},Z^n|Q^n,\mathcal{C})-H(Z^n|Q^n,\mathcal{C})\\ 
 			= 	
 			& H(M_\mathcal{K}, M_{\mathcal{K}, r}|Q^n,\mathcal{C})+H(Z^n|M_\mathcal{K}, M_{\mathcal{K}, r},Q^n,\mathcal{C})\\
 			&-H(Z^n|Q^n,\mathcal{C})-H(M_{\mathcal{K}, r}|M_\mathcal{K}, Z^n,Q^n,\mathcal{C})\\
 			\stackrel{(a)}{=} 	
 			& H(M_\mathcal{K}, M_{\mathcal{K}, r}|Q^n,\mathcal{C})+H(Z^n|V_1^n, \cdots, V_{K}^n, M_\mathcal{K}, M_{\mathcal{K}, r},Q^n,\mathcal{C})\\
 			&-H(Z^n|Q^n,\mathcal{C})-H(M_{\mathcal{K},r}|M_\mathcal{K},Z^n,Q^n,\mathcal{C})\\
 			\stackrel{(b)}{=} 
 			& n\sum_{i\in \mathcal{K}}[R_i+R_{i,r}]+H(Z^n|V_1^n, V_2^n, \cdots, V_{K}^n,Q^n,\mathcal{C})\\
 			&-H(Z^n|Q^n,\mathcal{C})-H(M_{\mathcal{K},r}|M_{\mathcal{K}}, Z^n,Q^n,\mathcal{C})\\
 			\stackrel{(c)}{\geq} 
 			& n\sum_{i\in \mathcal{K}}[R_i+R_{i,r}]-I(V_1^n, \cdots, V_{K}^n;Z^n|Q^n,\mathcal{C})\\
 			& -n\left[\sum_{i\in \mathcal{K}}R_{i,r}-I(V_{\mathcal{K}};Z|Q)+\varepsilon_n\right]
 			\stackrel{(d)}{\geq} 
 			 n\sum_{i\in \mathcal{K}}R_i-n\tau_n,
 	\end{align*}
 where $(a)$ follows from the fact that $V_1^n, \cdots, V_{K}^n$ are functions of $(M_{1}, M_{1,r})$, $\cdots$, $(M_{K}, M_{K,r}),$ respectively, given $Q^n$ and $\mathcal{C}$; $(b)$ follows from the fact that $H(M_\mathcal{K}, M_{\mathcal{K}, r}|Q^n,\mathcal{C})=n\sum\limits_{i\in \mathcal{K}}[R_i+R_{i,r}],$ and given $Q^n$ and $\mathcal{C},$ $(M_\mathcal{K}, M_{\mathcal{K}, r})\to (V_1^n, V_2^n, \cdots, V_{K}^n)\to Z^n$ forms a Markov chain; $(c)$ follows from  
 Lemma \ref{lem: Equivocation at Eve} (with $R_{v,i}=R_{i,r}$ for $i\in \mathcal{K}$) by requiring  \eqref{eqn:Lem Rates constraints on equivocation}: 
\begin{equation} \label{eqn: conds on s-collective secrecy}
\sum\limits_{j\in \mathcal{J}}R_{j,r}\geq I(V_{\mathcal{J}};Z|Q), \quad \forall \mathcal{J}\subseteq \mathcal{K}. 
\end{equation}
 		$(d)$ is due to the fact that $I(V_1^n,  \cdots, V_K^n; Z^n|Q^n,\mathcal{C})\leq n[I(V_1, \cdots, V_K;Z|Q)+\varepsilon_n],$ the proof of which follows the proof of \cite[Lemma 3]{src:Liu2008}, and taking $\tau_n=2\varepsilon_n.$
		
{\em Joint secrecy achievable rate region:} We summarize the requirements in order to guarantee a reliable communication under the joint secrecy constraint as follows:
\begin{itemize}
		\item the non-negativity for rates;
		\item the conditions for a reliable communication, i.e., \eqref{eqn:RateConstraints}; 
		\item the conditions for the joint secrecy, i.e., \eqref{eqn: conds on s-collective secrecy}.
\end{itemize}		
That is, we have the following system of inequalities:
 				\begin{align*}
					 \sum_{j\in \mathcal{J}} [R_{j}+R_{j,r}] \leq & I(V_{\mathcal{J}};Y|V_{\mathcal{J}^c},Q), \quad \mbox{for all}\ \mathcal{J}\subseteq \mathcal{K},\\
					\sum\limits_{j\in \mathcal{J}}R_{j,r}\geq & I(V_{\mathcal{J}};Z|Q), \quad \mbox{for all}\  \mathcal{J}\subseteq \mathcal{K}.						
 				\end{align*}
To obtain the desired region of $\{R_i|i\in \mathcal{K}\},$ the variables of $\{R_{i, r}|i\in \mathcal{K}\}$ are to be eliminated. (The rate constraints $R_i\geq 0$ for $i\in \mathcal{K}$ are not included here, since they will not be involved in the Fourier-Motzkin elimination. But they will be included in the final derived region.)  

{\em Fourier-Motzkin Elimination:}
A matrix notation of the above system of inequalities can be written as follows:
{\small
\begin{equation} \label{FM: system 1}
	\underbrace{
	\left[ 
	\begin{array}{c}
		\left[\mathrm{1}_\mathcal{J}\right]\\
		\\
		\left[\mathrm{0}_\mathcal{J}\right]
	\end{array} 
	\right]
	}_{\mathrm{A}'}
		 \cdot 
	\underbrace{
	\left[ 
	\begin{array}{c}
		R_1\\
		\vdots\\
		R_{K}
	\end{array} 
	\right]	
	}_{\mathrm{x}'}	 
	+
		\underbrace{
		\left[ 
		\begin{array}{r}
			\left[\mathrm{1}_\mathcal{J}\right]\\
			\\
			-\left[\mathrm{1}_\mathcal{J}\right]
		\end{array} 
		\right]
			}_{\mathrm{A}''}
			 \cdot
	 \underbrace{ 
		\left[ 
		\begin{array}{c}
			R_{1,r}\\
			\vdots\\
			R_{K,r}
		\end{array} 
		\right]
		}_{\mathrm{x}''}	 	
		\leq
			 \underbrace{ 
				\left[ 
				\begin{array}{r}
					\left[b^+_\mathcal{J}\right]\\
					\\
					-\left[b^-_\mathcal{J}\right]
				\end{array} 
				\right]
		}_{\mathrm{b}},
\end{equation}
}
where 
\begin{itemize}
	\item $\mathrm{1}_\mathcal{J}$ is the $1\times K$ indicator vector of the subset $\mathcal{J}$ of the set $\mathcal{K}$ (which definition is given in Definition \ref{app: def indicator vector}); 
	\item 
	Since $\mathcal{J}=\emptyset$ introduces only redundant inequations, only the non-empty choices of $\mathcal{J}$ need to be considered. Let $\{\mathcal{J}_1, \cdots, \mathcal{J}_{2^K-1}\}$ be the list of $2^K-1$ different non-empty subsets of $\mathcal{K}$. We have $\mathcal{J}\in \{\mathcal{J}_1, \cdots, \mathcal{J}_{2^K-1}\};$ 
	\item $\left[\mathrm{1}_\mathcal{J}\right]$ is a $(2^K-1)\times K$ matrix, with the $i$-th row to be the indicator vector $\mathrm{1}_{\mathcal{J}_i}$, where $1\leq i\leq 2^K-1.$ Correspondingly,
	\begin{itemize}
		\item  $\left[b^+_{\mathcal{J}}\right]$ is a $(2^K-1)\times 1$ matrix with the $i$-th row to be $b^+_{\mathcal{J}_i}=I(V_{\mathcal{J}_i};Y|V_{\mathcal{J}_i^c},Q),$
		\item  $\left[b^-_{\mathcal{J}}\right]$ is a $(2^K-1)\times 1$ matrix with the $i$-th row to be $b^-_{\mathcal{J}_i}=I(V_{\mathcal{J}_i};Z|Q)$;
	\end{itemize}	 
	\item $\left[\mathrm{0}_\mathcal{J}\right]$ is a $(2^K-1)\times K$ matrix with zero elements.
\end{itemize}
Properties of $b^+_{\mathcal{J}}$ and $b^-_{\mathcal{J}}$ are given in Lemma \ref{app: lem for t T} in Appendix \ref{App: Compact form}, which play a crucial role in  removing the redundant inquations so as to avoid the double exponential complexity of the Fourier-Motzkin elimination.

Note that $\mathrm{A}', \mathrm{A}''$ are both $(2^{K+1}-2)\times K$ matrices. 
In particular, we have 
\begin{align}
\allowdisplaybreaks
\mathrm{A}''
		&=	\left[ 
				\begin{array}{r}
					\left[\mathrm{1}_\mathcal{J}\right]\\
					-\left[\mathrm{1}_\mathcal{J}\right]
				\end{array} 
				\right]
		={\mathrm{B}} \otimes \left[\mathrm{1}_\mathcal{J}\right],
		\ 
		\mbox{where}\ 
		\mathrm{B}=\left[
					\begin{array}{r}
						1\\
						-1
					\end{array}
				\right].
		 \label{FM: A'' representation}
\end{align}
Here $\otimes$ is the Kronecker product.  
According to \cite[Theorem 1]{src:Imbert93fm}, we are looking for a $G$ such that $\{\mathrm{w} \mathrm{A}' \mathrm{x}' \leq \mathrm{w} \mathrm{b}| \mathrm{w}\in \mathrm{G}\}$ is equivalent to the final system of the Fourier-Motzkin elimination. Here, $\mathrm{G}$ is a base of essentially different minimal vectors of $\mathrm{F}$ that is the cone of non-negative solution $\mathrm{w}\geq 0$ of $\mathrm{w}\mathrm{A}''=0$. 

In fact, for our system defined by \eqref{FM: system 1}, we find 
\begin{equation}\label{eqn: matrix G}
	\mathrm{G}=\mathrm{D}\otimes \mathrm{I}_{2^K-1},\quad \mbox{where}\ \mathrm{D}=\left[
					\begin{array}{cc}
						1 & 1
					\end{array}
				\right].
\end{equation}
Here $\mathrm{I}_n$ is an $n\times n$ identity matrix. 
In the following, we show that $\mathrm{G}$ (as defined in \eqref{eqn: matrix G}) is a base of essentially different minimal vectors of $\mathrm{F}.$ Note that the system defined by $\{\mathrm{w}\mathrm{A}'\mathrm{x}'\leq \mathrm{w}\mathrm{b}|\mathrm{w}\in \mathrm{G}\}$ is:  
 \begin{equation}\label{eqn: final system}
 	\mathrm{G}\mathrm{A}'\mathrm{x}'\leq  \mathrm{G}\mathrm{b} \quad \Leftrightarrow \quad
 	\left[\mathrm{1}_\mathcal{J}\right] \mathrm{x}'\leq  \left[{b}^+_\mathcal{J}-{b}^-_\mathcal{J}\right].
 \end{equation}
 
Firstly, it is clear that $\mathrm{w}\in \mathrm{G}$ is positive, i.e., $\mathrm{w}\geq \mathrm{0}.$  

Secondly, we show that $\mathrm{w}\mathrm{A}''=0$ for any $\mathrm{w}\in \mathrm{G}.$ Note that $\mathrm{A}''={\mathrm{B}} \otimes \left[\mathrm{1}_\mathcal{J}\right]$ by \eqref{FM: A'' representation}. Therefore, if $\mathrm{w}\cdot(\mathrm{B}\otimes \left[\mathrm{1}_\mathcal{J}\right])=0,$ we have $\mathrm{w}\cdot \mathrm{A}''=0$ as well. It is easy to verify that 
\begin{equation*}
	\mathrm{G}\cdot({\mathrm{B}} \otimes \left[\mathrm{1}_\mathcal{J}\right])=(\mathrm{D}\otimes \mathrm{I}_{2^K-1}) \cdot({\mathrm{B}} \otimes \left[\mathrm{1}_\mathcal{J}\right]) \stackrel{(a)}{=} (\mathrm{D}\mathrm{B})\otimes \left[\mathrm{1}_\mathcal{J}\right]=\left[\mathrm{0}_\mathcal{J}\right],
\end{equation*}
where $(a)$ is due to the mixed-product property of the Kronecker product: $(\mathrm{A}\otimes \mathrm{B})(\mathrm{C}\otimes \mathrm{D})=(\mathrm{A}\mathrm{C})\otimes (\mathrm{B}\mathrm{D}).$

Furthermore, by definition of $\mathrm{G}$ in \eqref{eqn: matrix G}, each row of $\mathrm{G}$ is essentially different from the others. And, they are all minimal according to \cite[Theorem 2]{src:Imbert93fm}.  

Most importantly, we show in the following that any other positive vectors $\mathrm{c}$ satisfying $\mathrm{c}\cdot \mathrm{A}''=0$, will produce only redundant inequations. 
For convenience, we denote $\mathrm{c}=[\mathrm{c}^+ \ \mathrm{c}^-],$ with $\mathrm{c}^+=[c^+_1, \cdots, c^+_{2^K-1}]$ and $\mathrm{c}^-=[c^-_1, \cdots, c^-_{2^K-1}]$. Then, $\mathrm{c}\cdot \mathrm{A}''=0$ is equivalent to $\mathrm{c}^+\cdot \left[\mathrm{1}_\mathcal{J}\right]=\mathrm{c}^-\cdot \left[\mathrm{1}_\mathcal{J}\right].$ 

Recall that for $1\leq i\leq 2^K-1,$ $\mathrm{1}_{\mathcal{J}_i}$ is the indicator vector of $\mathcal{J}_i,$ and thus $c^+_{i}>0$ (or $c^-_{i}>0$) indicates $c^+_{i}$ (or $c^-_{i}$) number of presences of $\mathcal{J}_i.$ Let
\begin{align*}
 J_{\mathrm{c}^+}=& \{\underbrace{\mathcal{J}_1, \cdots, \mathcal{J}_1}_{c^+_{1}}, \cdots,\underbrace{\mathcal{J}_i, \cdots, \mathcal{J}_i}_{c^+_{i}},  \cdots, \underbrace{\mathcal{J}_{2^K-1}, \cdots, \mathcal{J}_{2^K-1}}_{c^+_{2^K-1}}\};\\
 J_{\mathrm{c}^-}=& \{\underbrace{\mathcal{J}_1, \cdots, \mathcal{J}_1}_{c^-_{1}}, \cdots,\underbrace{\mathcal{J}_i, \cdots, \mathcal{J}_i}_{c^-_{i}},  \cdots, \underbrace{\mathcal{J}_{2^K-1}, \cdots, \mathcal{J}_{2^K-1}}_{c^-_{2^K-1}}\}.
\end{align*} 
Denote $n^+=\sum\limits_{i=1}^{2^K-1} c^+_{i}$ and ${n^-}=\sum\limits_{i=1}^{2^K-1} c^-_{i}.$ 
Simply representing $J_{\mathrm{c}^+}=\{\mathcal{J}^+_1, \mathcal{J}^+_2, \cdots, \mathcal{J}^+_{n^+}\}$ and $J_{\mathrm{c}^-}=\{\mathcal{J}^-_1, \mathcal{J}^-_2, \cdots, \mathcal{J}^-_{n^-}\}$, we have
\begin{align}
	\mathrm{c}^+\cdot \left[\mathrm{1}_\mathcal{J}\right]=\sum\limits_{i=1}^{n^+} \mathrm{1}_{\mathcal{J}_i^+},\quad & 
	\mathrm{c}^-\cdot \left[\mathrm{1}_\mathcal{J}\right]=\sum\limits_{i=1}^{n^-} \mathrm{1}_{\mathcal{J}_i^-};\label{eqn: c+1J}\\
	\mathrm{c}^+\cdot \left[{b}^+_\mathcal{J}\right]=\sum_{i=1}^{n^+} {b}^+_{\mathcal{J}_i^+}, \quad & \mathrm{c}^-\cdot \left[{b}^-_\mathcal{J}\right]=\sum_{i=1}^{n^-} {b}^-_{\mathcal{J}_i^-}. \label{eqn: c+bJ}
\end{align}
Note that $\mathrm{c}^+\cdot \left[\mathrm{1}_\mathcal{J}\right]=\mathrm{c}^-\cdot \left[\mathrm{1}_\mathcal{J}\right]$ and \eqref{eqn: c+1J} together 
imply that both $J_{\mathrm{c}^+}$ and $J_{\mathrm{c}^-}$ 
share the same presence vector $\mathrm{c}_{\#}=\mathrm{c}^+\cdot \left[\mathrm{1}_\mathcal{J}\right]$ (according to Definition \ref{app: def presence vector}). Denote the largest element of $\mathrm{c}_{\#}$ to be $c_{\max}.$ 

Let  $\{\mathcal{J}^{\oplus}_{i} |1\leq i\leq {n^+}\}$ and $\{\mathcal{J}^{\ominus}_{i} |1\leq i\leq {n^-}\}$ be {\em the compact forms of the element rearrangement} (definition of which is given in Definition \ref{app: def compact form}) for  $J_{\mathrm{c}^+}$ and $J_{\mathrm{c}^-},$  
respectively. Then, according to Lemma \ref{lem: app compact arragement properties}, they also share the same {\em compact form of the element rearrangement without empty sets}, i.e., $ \{\mathcal{J}^{\oplus}_{i} |1\leq i\leq c_{\max}\}.$ That is,
\begin{align}
	\mathcal{J}^{\oplus}_{i} =\mathcal{J}^{\ominus}_{i}, &\quad \mbox{for}\quad 1\leq i\leq c_{\max}; \label{eqn: Jn+=Jn-}\\ 
	\mathcal{J}^{\oplus}_{i} =\mathcal{J}^{\ominus}_{j} =\emptyset, &\quad \mbox{for}\quad  c_{\max}<i\leq n^+ \ \& \ c_{\max}<j\leq n^-.\nonumber
\end{align}
Since the element arrangement does not change the presence vector, we have by Lemma \ref{lem: invariability of presence vector}:
\begin{equation}\label{eqn: LHS cAx'}
	\mathrm{c}_{\#}=\mathrm{c}^+\cdot \left[\mathrm{1}_\mathcal{J}\right]=\sum_{i=1}^{c_{\max}} \mathrm{1}_{\mathcal{J}^{\oplus}_{i}}.
\end{equation}

Now if we sum up the rows of \eqref{eqn: final system} with respect to $\{\mathcal{J}^{\oplus}_{i} |1\leq i\leq  c_{\max}\},$ we obtain the following inequation:
\begin{equation}\label{eqn: linear combination of final system}
	\sum_{i=1}^{c_{\max}} \mathrm{1}_{\mathcal{J}^{\oplus}_{i}} \mathrm{x}' \leq \sum_{i=1}^{c_{\max}} \left[{b}^+_{\mathcal{J}^{\oplus}_{i}}-{b}^-_{\mathcal{J}^{\oplus}_{i}}\right].
\end{equation}
On the other hand, if we apply $\mathrm{c}$ to the original system \eqref{FM: system 1}, we obtain the following inequation:
\begin{align}
	\mathrm{c}\mathrm{A}'\mathrm{x}'\leq \mathrm{c}\mathrm{b}  
	\quad &\Rightarrow \quad \mathrm{c}^+ \left[\mathrm{1}_\mathcal{J}\right] \mathrm{x}'\leq  \mathrm{c}^+\left[{b}^+_\mathcal{J}\right]-\mathrm{c}^-\left[{b}^-_\mathcal{J}\right] \nonumber\\
	\quad &\stackrel{(a)}{\Rightarrow} \quad \mathrm{c}_{\#} \mathrm{x}'\leq  \sum_{i=1}^{n^+} {b}^+_{\mathcal{J}_i^+}- \sum_{i=1}^{n^-} {b}^-_{\mathcal{J}_i^-}, \label{eqn: cAx'}
\end{align}
where $(a)$ is due to \eqref{eqn: LHS cAx'} and \eqref{eqn: c+bJ}. 

Note that the LHS of \eqref{eqn: cAx'} is the same as the LHS of \eqref{eqn: linear combination of final system} (according to \eqref{eqn: LHS cAx'}). 
However, we can show that 
\begin{align*}
	\mbox{RHS of \eqref{eqn: cAx'}} =& 
	\sum_{i=1}^{n^+} {b}^+_{\mathcal{J}_i^+}- \sum_{i=1}^{n^-} {b}^-_{\mathcal{J}_i^-}
	\stackrel{(b)}{\geq}  \sum_{i=1}^{n^+} {b}^+_{\mathcal{J}^{\oplus}_{i}}- \sum_{i=1}^{n^-} {b}^-_{\mathcal{J}^{\ominus}_{i} }\\
	\stackrel{(c)}{=}&  \sum_{i=1}^{c_{\max}} \left[{b}^+_{\mathcal{J}^{\oplus}_{i}}-{b}^-_{\mathcal{J}^{\oplus}_{i}}\right]
	= 	\mbox{RHS of \eqref{eqn: linear combination of final system}},
\end{align*}
where $(b)$ is according to Lemma \ref{app: lem for t T} (as given in Appendix \ref{App: Compact form}) and $(c)$ is due to \eqref{eqn: Jn+=Jn-} and $b^+_{\emptyset}=b^-_{\emptyset}=\mathrm{0}$. 
That is, \eqref{eqn: cAx'} is redundant since it is already implied by \eqref{eqn: linear combination of final system}, which can be derived as a linear combinatory with positive coefficients of inequations of \eqref{eqn: final system}. Since this applies to any positive vector $\mathrm{c}$ that is other than the rows of $\mathrm{G}$ (as defined in \eqref{eqn: matrix G}) such that $\mathrm{c}\cdot \mathrm{A}''=0,$ therefore, the final system of \eqref{FM: system 1} is equivalent to the one  \eqref{eqn: final system}. 

As a conclusion, \eqref{eqn: final system} establishes the resulting joint secrecy region. 
\end{IEEEproof}

\appendices

\section{Properties of $b^+_{\mathcal{T}}$ and $b^-_{\mathcal{T}}$}
\label{App: Compact form}
Recall that $b^+_{\mathcal
{T}}=I(V_{\mathcal{T}};Y|V_{\mathcal{T}^c},Q)$ and $b^-_{\mathcal{T}}=I(V_{\mathcal{T}};Z|Q)$ as defined in \eqref{FM: system 1}. 
We have the following lemma:
\begin{lemma} \label{app: lem for t T}
Given  $\{\mathcal{T}_i|1\leq i\leq t\}$ as a list of $t$ subsets of $\mathcal{K},$ and its compact form of the element rearrangement $\{\mathcal{T}_{t,i}^{\ast}|1\leq i\leq t\},$ we have
$$\sum\limits_{i=1}^{t} b^+_{\mathcal{T}_i}\geq \sum\limits_{i=1}^{t} b^+_{\mathcal{T}^{\ast}_{t,i}} \quad \mbox{and} \quad\sum\limits_{i=1}^{t} b^-_{\mathcal{T}_i}\leq \sum\limits_{i=1}^{t} b^-_{\mathcal{T}^{\ast}_{t,i}}.$$
\end{lemma}


To prove Lemma \ref{app: lem for t T}, we need the following two lemmas.
\begin{lemma} \label{lem: recursive from t-1 to t}
Given $\{\mathcal{T}_{i} |1\leq i\leq t-1\}$ as a list of $t-1$ subsets of $\mathcal{K},$ and list $\{\mathcal{T}_{i} |1\leq i\leq t\}$ (with one more subset $\mathcal{T}_t$ included),  let $\{\mathcal{T}_{t-1,i}^{\ast}|1\leq i\leq t-1\}$ and $\{\mathcal{T}_{t,i}^{\ast}|1\leq i\leq t\}$ be their compact forms of the element rearrangement, respectively. We have 
\begin{equation*}
\mathcal{T}_{t,i}^{\ast}=\left\{
\begin{array}{lll}
	 \mathcal{T}^{\ast}_{t-1,1}\cup \mathcal{T}_t & & i=1\\
	 \mathcal{T}^{\ast}_{t-1,i} \bigcup  \left(\mathcal{T}^{\ast}_{t-1,i-1}\bigcap
	 												\mathcal{T}_t\right)
	 												&& 1<i<t\\
	 \mathcal{T}^{\ast}_{t-1,t-1}\cap \mathcal{T}_t & & i=t\\												
 \end{array}
\right..
\end{equation*}
\end{lemma}
\begin{IEEEproof}
	\begin{table*}[ht]
	\vspace{-2mm}
	\begin{align}
		\mathcal{T}_{t,i}^{\ast} 
			= \bigcup_{\{j_1, \cdots, j_i\}\subseteq [1:t]} \left(\bigcap_{k=1}^i \mathcal{T}_{j_k}\right) 
			= & \left\{\bigcup_{\{j_1, \cdots, j_i\}\subseteq [1:t-1]} \left(\bigcap_{k=1}^i \mathcal{T}_{j_k}\right)\right\} 
			\bigcup \left\{\bigcup_{\{j_1, \cdots, j_{i-1}\}\subseteq [1:t-1]} \left(\left(\bigcap_{k=1}^{i-1} \mathcal{T}_{j_k}\right)\bigcap
			\mathcal{T}_t\right)\right\} \label{step1}\\
			{=}& 
			\left\{\bigcup_{\{j_1, \cdots, j_i\}\subseteq [1:t-1]} \left(\bigcap_{k=1}^i \mathcal{T}_{j_k}\right)\right\}
			\bigcup \left\{ \left(\bigcup_{\{j_1, \cdots, j_{i-1}\}\subseteq [1:t-1]} \left(\bigcap_{k=1}^{i-1} \mathcal{T}_{j_k}\right)\right)\bigcap
						\mathcal{T}_t\right\} \label{step2}\\
			{=}& \mathcal{T}^{\ast}_{t-1, i} \bigcup  \left(\mathcal{T}^{\ast}_{t-1,i-1}\bigcap
												\mathcal{T}_t\right),  \label{step3}
	\end{align}
	\end{table*}
The proof for $i=1,t$ is straightforward by definitions of $\mathcal{T}^{\ast}_{t,1}$ and $\mathcal{T}^{\ast}_{t,t}$. For $1<i<t,$ the proof is given in steps \eqref{step1}-\eqref{step3} (at the top of the next page), where \eqref{step1} is by the definition of $\mathcal{T}^{\ast}_{t,i}$; \eqref{step2} is due to the distributive laws of the sets and \eqref{step3} is by the definition of $\mathcal{T}^{\ast}_{t-1,i}.$
\end{IEEEproof}

\begin{lemma} \label{lem: b+ b- for t=2}
For $\forall \mathcal{T}_1, \mathcal{T}_2\subseteq \mathcal{K},$ we have
	\begin{enumerate}
		\item $b^+_{\mathcal{T}_1}+b^+_{\mathcal{T}_2}\geq b^+_{\mathcal{T}_1\cap \mathcal{T}_2}+b^+_{\mathcal{T}_1\cup \mathcal{T}_2}$;
		\item $b^-_{\mathcal{T}_1}+b^-_{\mathcal{T}_2}\leq b^-_{\mathcal{T}_1\cap \mathcal{T}_2}+b^-_{\mathcal{T}_1\cup \mathcal{T}_2}$.
	\end{enumerate}
\end{lemma}

\begin{IEEEproof}
	We only give the proof for $b^+_{\mathcal{T}_1}+b^+_{\mathcal{T}_2}.$ A similar proof applies to  $b^-_{\mathcal{T}_1}+b^-_{\mathcal{T}_2}.$
	\begin{align*}
		& b^+_{\mathcal{T}_1}+b^+_{\mathcal{T}_2} \stackrel{(a)}{=}I(V_{\mathcal{T}_1};Y|V_{{\mathcal{T}_1}^c},Q)+I(V_{\mathcal{T}_2};Y|V_{{\mathcal{T}_2}^c},Q)\\
		&\stackrel{(b)}{=}I(V_{\mathcal{T}_1\cap \mathcal{T}_2}, V_{\mathcal{T}_1\cap \mathcal{T}_2^c};Y|V_{{\mathcal{T}_1}^c},Q)+I(V_{\mathcal{T}_2};Y|V_{{\mathcal{T}_2}^c},Q)\\
		&\stackrel{(c)}{=}I(V_{\mathcal{T}_1\cap \mathcal{T}_2^c};Y|V_{{\mathcal{T}_1}^c},Q)+I(V_{\mathcal{T}_1\cap \mathcal{T}_2};Y|V_{{\left(\mathcal{T}_1\cap \mathcal{T}_2\right)}^c},Q)\\
		&+I(V_{\mathcal{T}_2};Y|V_{{\mathcal{T}_2}^c},Q)\\
		&\stackrel{(g)}{\geq} I(V_{\mathcal{T}_1\cap \mathcal{T}_2};Y|V_{{\left(\mathcal{T}_1\cap \mathcal{T}_2\right)}^c},Q)+I(V_{\mathcal{T}_1\cap \mathcal{T}_2^c};Y|V_{{\mathcal{T}_1}^c\cap{\mathcal{T}_2}^c },Q)\\
		&+I(V_{\mathcal{T}_2};Y|V_{{\mathcal{T}_1}\cap{\mathcal{T}_2}^c}, V_{{\mathcal{T}_1}^c\cap{\mathcal{T}_2}^c},Q)\\
		&\stackrel{(e)}{=}I(V_{\mathcal{T}_1\cap \mathcal{T}_2};Y|V_{{\left(\mathcal{T}_1\cap \mathcal{T}_2\right)}^c},Q)+I(V_{\mathcal{T}_1\cup \mathcal{T}_2};Y|V_{({\mathcal{T}_1}\cup{\mathcal{T}_2})^c},Q),
	\end{align*}
where $(a)$ is by the definition of $b^+_{\mathcal
{T}};$ $(b)$ is by the fact that $\mathcal{T}_1=({\mathcal{T}_1\cap \mathcal{T}_2})\cup({\mathcal{T}_1\cap \mathcal{T}_2^c});$ $(c)$ is by the chain rule of the mutual information; $(d)$ is by the facts that $\mathcal{T}_2^c=({\mathcal{T}_1\cap \mathcal{T}_2^c})\cup({\mathcal{T}_1^c\cap \mathcal{T}_2^c})$ and  $I(V_{\mathcal{T}_1\cap \mathcal{T}_2^c};Y|V_{{\mathcal{T}_1}^c},Q)=I(V_{\mathcal{T}_1\cap \mathcal{T}_2^c};V_{{\mathcal{T}_1}^c\cap{\mathcal{T}_2}},Y|V_{{\mathcal{T}_1}^c\cap{\mathcal{T}_2}^c },Q)\geq I(V_{\mathcal{T}_1\cap \mathcal{T}_2^c};Y|V_{{\mathcal{T}_1}^c\cap{\mathcal{T}_2}^c},\allowbreak Q)$, which holds since $V_{\mathcal{T}_1\cap \mathcal{T}_2^c},$ $V_{{\mathcal{T}_1}^c\cap{\mathcal{T}_2}}$ and $V_{{\mathcal{T}_1}^c\cap{\mathcal{T}_2}^c}$ are independent  given $Q$; and $(e)$ is due to the facts that $\mathcal{T}_1\cup \mathcal{T}_2={\mathcal{T}_2}\cup({\mathcal{T}_1\cap \mathcal{T}_2^c})$ and ${\mathcal{T}_1}^c\cap{\mathcal{T}_2}^c={({\mathcal{T}_1}\cup{\mathcal{T}_2})^c}.$
\end{IEEEproof}

Now we give the proof of Lemma \ref{app: lem for t T} as follows.

First, we show that the statement $\sum\limits_{i=1}^{t} b^+_{\mathcal{T}_i}\geq \sum\limits_{i=1}^{t} b^+_{\mathcal{T}^{\ast}_{t,i}}$ is true by induction. 
\begin{itemize}
	\item 	For $t=1,$ we have by definition $\mathcal{T}^{\ast}_{1,1}=\mathcal{T}_1.$ Thus,  $b^+_{\mathcal{T}_1}=b^+_{\mathcal{T}^{\ast}_{1,1}}$  and the statement is true for $t=1.$
	\item 	
		For  $t=2$, the statement is true by Lemma \ref{lem: b+ b- for t=2}. 
	\item 	Assume that the statement is true for some natural number $t-1.$ That is, for any $\{\mathcal{T}_i|1\leq i\leq t-1\}$ as a list of $t-1$ subsets of $\mathcal{K},$ and its compact form of element rearrangement $\{\mathcal{T}_{t-1,i}^{\ast}|1\leq i\leq t-1\},$ we have
		$\sum\limits_{i=1}^{t-1} b^+_{\mathcal{T}_i} \geq \sum\limits_{i=1}^{t-1} b^+_{\mathcal{T}_{t-1,i}^{\ast}}.$
	\item 	Now we show that the statement is also true for $t$. 
		\begin{align*}
			\sum_{i=1}^{t} b^+_{\mathcal{T}_i} 
				 = &  \sum_{i=1}^{t-1} b^+_{\mathcal{T}_i} + b^+_{\mathcal{T}_t}
				 \stackrel{(a)}{\geq}  \sum_{i=1}^{t-1} b^+_{\mathcal{T}^{\ast}_{t-1,i}} + b^+_{\mathcal{T}_t}\\
				=& \sum_{i=2}^{t-1} b^+_{\mathcal{T}^{\ast}_{t-1,i}}+\left(b^+_{\mathcal{T}^{\ast}_{t-1,1}} + b^+_{\mathcal{T}_t}\right)\\
				 \stackrel{(b_1)}{\geq}&  \sum_{i=2}^{t-1} b^+_{\mathcal{T}^{\ast}_{t-1,i}}+\left(b^+_{{\mathcal{T}^{\ast}_{t-1,1}}\cap\mathcal{T}_t}+ b^+_{{\mathcal{T}^{\ast}_{t-1,1}}\cup\mathcal{T}_t}\right)\\
				 \stackrel{(c_1)}{=}&  \sum_{i=3}^{t-1} b^+_{\mathcal{T}^{\ast}_{t-1,i}}+\left(b^+_{\mathcal{T}^{\ast}_{t-1,2}}+b^+_{{\mathcal{T}^{\ast}_{t-1,1}}\cap\mathcal{T}_t}\right)+ b^+_{\mathcal{T}^{\ast}_{t,1}}\\	
				 \stackrel{(b_2)}{\geq}&  \sum_{i=3}^{t-1} b^+_{\mathcal{T}^{\ast}_{t-1,i}}+\left(b^+_{\mathcal{T}^{\ast}_{t-1,2}\cap\mathcal{T}_t}+b^+_{\mathcal{T}^{\ast}_{t-1,2}\cup\left({\mathcal{T}^{\ast}_{t-1,1}}\cap\mathcal{T}_t\right)}\right)\\
				 &+ b^+_{\mathcal{T}^{\ast}_{t,1}}\\	
				 \stackrel{(c_2)}{=} & \sum_{i=4}^{t-1} b^+_{\mathcal{T}^{\ast}_{t-1,i}}+\left(b^+_{\mathcal{T}^{\ast}_{t-1,3}}+b^+_{\mathcal{T}^{\ast}_{t-1,2}\cap\mathcal{T}_t}\right)+  \sum_{i=1}^{2}b^+_{\mathcal{T}^{\ast}_{t,i}}\\	
				& \vdots\\
				 \stackrel{(b_{t-1})}{\geq} & b^+_{\mathcal{T}^{\ast}_{t-1,t-1}\cap\mathcal{T}_t}+b^+_{\mathcal{T}^{\ast}_{t-1,t-1}\cup\left({\mathcal{T}^{\ast}_{t-1,t-2}}\cap\mathcal{T}_t\right)}+ \sum_{i=1}^{t-2}b^+_{\mathcal{T}^{\ast}_{t,i}}\\	
				 \stackrel{(c_{t-1})}{=}&   \sum_{i=1}^{t}b^+_{\mathcal{T}^{\ast}_{t,i}},	
		\end{align*}
	where $(a)$ is due to the fact that the statement is true for $t-1;$ and for $1\leq j\leq t-1,$ step $(b_j)$ is by applying Lemma \ref{lem: b+ b- for t=2} and the fact that $\mathcal{T}^{\ast}_{t-1,j}\cap \left({\mathcal{T}^{\ast}_{t-1,j-1}}\cap\mathcal{T}_t\right)=\mathcal{T}^{\ast}_{t-1,j}\cap \mathcal{T}_t$ for $2\leq j\leq t-1$ (since $\mathcal{T}^{\ast}_{t-1,j}\subseteq \mathcal{T}^{\ast}_{t-1,j-1}$ by definition);  step $(c_j)$ is by applying Lemma \ref{lem: recursive from t-1 to t}. In particular, $(c_1)$ is by the fact that $\mathcal{T}^{\ast}_{t,1} =\mathcal{T}^{\ast}_{t-1,1} \bigcup \mathcal{T}_t;$ step  $(c_{t-1})$ is by the fact that $\mathcal{T}^{\ast}_{t,t} =\mathcal{T}^{\ast}_{t-1,t-1} \bigcap \mathcal{T}_t;$ and other intermediate steps are by the fact that $\mathcal{T}^{\ast}_{t-1,j}\cup \left({\mathcal{T}^{\ast}_{t-1,j-1}}\cap\mathcal{T}_t\right)=\mathcal{T}^{\ast}_{t,j}$ for $1<j<t.$
\end{itemize}	

A similar proof can be applied to show that the statement for $\sum\limits_{i=1}^{t} b^-_{\mathcal{T}_i}$ is also true. This concludes our proof. 	

\end{document}